\documentclass{aa}
\usepackage{graphicx}
\begin{document}

\title{The WEBT \object{BL Lacertae} Campaign 2001 and its 
extension}

\subtitle{Optical light curves and colour analysis 1994--2002}

\author{M.~Villata \inst{1}
\and C.~M.~Raiteri \inst{1}
\and O.~M.~Kurtanidze \inst{2,3,4}
\and M.~G.~Nikolashvili \inst{2}
\and M.~A.~Ibrahimov \inst{5}
\and I.~E.~Papadakis \inst{6,7}
\and G.~Tosti \inst{8}
\and F.~Hroch \inst{9}
\and L.~O.~Takalo \inst{10}
\and A.~Sillanp\"a\"a \inst{10}
\and V.~A.~Hagen-Thorn \inst{11,12}
\and V.~M.~Larionov \inst{11,12}
\and R.~D.~Schwartz \inst{13}
\and J.~Basler \inst{13}
\and L.~F.~Brown \inst{14}
\and T.~J.~Balonek \inst{15}
\and E.~Ben\'{\i}tez \inst{16}
\and A.~Ram\'{\i}rez \inst{16}
\and A.~C.~Sadun \inst{17}
\and P.~Boltwood \inst{18}
\and M.~T.~Carini \inst{19}
\and D.~Barnaby \inst{19}
\and J.~M.~Coloma \inst{20}
\and J.~A.~Ros \inst{20}
\and B.~Z.~Dai \inst{21,22,23}
\and G.~Z.~Xie \inst{21,22,23}
\and J.~R.~Mattox \inst{24}
\and D.~Rodriguez \inst{25}
\and I.~M.~Asfandiyarov \inst{5}
\and A.~Atkerson \inst{19} 
\and J.~L.~Beem \inst{14}
\and S.~D.~Bloom \inst{26}
\and S.~M.~Chanturiya \inst{2}
\and S.~Ciprini \inst{8}
\and S.~Crapanzano \inst{1}
\and J.~A.~de~Diego \inst{16}
\and N.~V.~Efimova \inst{11}
\and D.~Gardiol \inst{1,27}
\and J.~C.~Guerra \inst{27}
\and B.~B.~Kahharov \inst{5}
\and B.~Z.~Kapanadze \inst{2}
\and H.~Karttunen \inst{10}
\and T.~Kato \inst{28}
\and G.~N.~Kimeridze \inst{2}
\and N.~A.~Kudryavtseva \inst{11}
\and M.~Lainela \inst{10}
\and L.~Lanteri \inst{1}
\and E.~G.~Larionova \inst{11}
\and M.~Maesano \inst{29}
\and N.~Marchili \inst{8}
\and G.~Massone \inst{1}
\and T.~Monroe \inst{19}
\and F.~Montagni \inst{30}
\and R.~Nesci \inst{31}
\and K.~Nilsson \inst{10}
\and J.~C.~Noble \inst{32}
\and G.~Nucciarelli \inst{8}
\and L.~Ostorero \inst{4,33}
\and J.~Papamastorakis \inst{7,6}
\and M.~Pasanen \inst{10}
\and C.~S.~Peters \inst{14}
\and T.~Pursimo \inst{34}
\and P.~Reig \inst{35,6}
\and W.~Ryle \inst{19}
\and S.~Sclavi \inst{31}
\and L.~A.~Sigua \inst{2}
\and M.~Uemura \inst{28}
\and W.~Wills \inst{19}
} 

\offprints{M.\ Villata, \email{villata@to.astro.it}} 

\institute{Istituto Nazionale di Astrofisica (INAF), 
Osservatorio Astronomico di Torino, Via Osservatorio 20, 
10025 Pino Torinese (TO), Italy 
\and Abastumani Astrophysical Observatory, 383762 
Abastumani, Georgia
\and Astrophysikalisches Institut Potsdam, An der 
Sternwarte 16, 14482 Potsdam, Germany
\and Landessternwarte Heidelberg-K\"onigstuhl, 
K\"onigstuhl 12, 69117 Heidelberg, Germany
\and Ulugh Beg Astronomical Institute, Academy of 
Sciences of Uzbekistan, 33 Astronomical Str., Tashkent 
700052, Uzbekistan
\and IESL, FORTH, 711 10 Heraklion, Crete, Greece
\and Physics Department, University of Crete, PO Box 
2208, 710 03 Heraklion, Crete, Greece
\and Dipartimento di Fisica, Universit\`a di Perugia, 
Via A.\ Pascoli, 06123 Perugia, Italy
\and Institute of Theoretical Physics and Astrophysics, 
Faculty of Science, Masaryk University, 
Kotl\'{a}\v{r}sk\'{a} 2, 611 37 Brno, Czech Republic
\and Tuorla Observatory, 21500 Piikki\"o, Finland
\and Astronomical Institute, St.-Petersburg
State University, Universitetsky Pr.\ 28,
Petrodvoretz, 198504 St.-Petersburg, Russia
\and Isaac Newton Institute of Chile, St.-Petersburg 
Branch
\and Department of Physics and Astronomy, University 
of Missouri-St.\ Louis, 8001 Natural Bridge Road,
St.\ Louis, MO 63121, USA
\and Department of Physics, Astronomy and Geophysics,
Connecticut College, New London, CT 06320, USA
\and Foggy Bottom Observatory, Colgate University, 13 
Oak Drive, Hamilton, NY 13346, USA
\and Instituto de Astronom\'{\i}a, UNAM, 
Apdo.\ Postal 70-264, 04510 M\'exico DF, Mexico 
\and Department of Physics, University of Colorado 
at Denver, PO Box 173364, Denver, CO 80217-3364, USA
\and Boltwood Observatory, 1655 Main Street, 
Stittsville, Ontario K2S 1N6, Canada
\and Department of Physics and Astronomy, Western 
Kentucky University, 1 Big Red Way, Bowling Green, KY 
42104, USA
\and Agrupaci\'o Astron\`omica de Sabadell, PO Box 
50, 08200 Sabadell, Spain
\and Yunnan Observatory, National Astronomical 
Observatories, Chinese Academy of Sciences, PO Box 110, 
Kunming 650011, China 
\and United Laboratory of Optical Astronomy, Chinese 
Academy of Sciences, Beijing, China
\and Yunnan Astrophysics Center, Yunnan University, 
Kunming 650091, China
\and Department of Natural Sciences, Fayetteville State 
University, 1200 Murchison Road, Fayetteville, NC 28301,
USA
\and Guadarrama Observatory, C/ San Pablo 5, 
Villalba 28409, Madrid, Spain
\and Department of Physics and Astronomy, 
Hampden-Sydney College, Hampden-Sydney, VA 23943, USA
\and Istituto Nazionale di Astrofisica (INAF), 
Telescopio Nazionale Galileo, Roque de los 
Muchachos Astronomical Observatory, PO Box 565, 38700 
Santa Cruz de La Palma, TF, Spain
\and Department of Astronomy, Faculty of Science, 
Kyoto University, Kyoto, Japan
\and Stazione Astronomica Vallinfreda, Italy
\and Stazione Astronomica Greve in Chianti, Italy
\and Dipartimento di Fisica, Universit\`a La Sapienza,
Piazzale A.\ Moro 2, 00185 Roma, Italy
\and Institute for Astrophysical Research, Boston 
University, 725 Commonwealth Ave., Boston, MA 02215, USA
\and Dipartimento di Fisica Generale, Universit\`a di 
Torino, Via P.\ Giuria 1, 10125 Torino, Italy 
\and Nordic Optical Telescope, Roque de los 
Muchachos Astronomical Observatory, PO Box 474, 
38700 Santa Cruz de La Palma, TF, Spain 
\and G.A.C.E., Departament d'Astronomia i 
Astrof\'{\i}sica, Universitat de Val\`encia, 46071 
Paterna-Val\`encia, Spain
}

\date{Received; Accepted;}

\titlerunning{The WEBT BL Lac Campaign 2001 and its extension}

\authorrunning{M.\ Villata et al.}

\abstract{see next page
}

\maketitle

\twocolumn[{\bf Abstract.}
BL Lacertae has been the target of four observing campaigns by the Whole Earth Blazar 
Telescope (WEBT) collaboration. In this paper we present $UBVRI$ light curves obtained
by the WEBT from 1994 to 2002, including the last, extended BL Lac 2001 campaign. 
A total of about 7500 optical observations performed by 31 telescopes from Japan to Mexico
have been collected, to be added to the $\sim 15600$ observations of the BL Lac Campaign 2000. 
All these data allow one to follow the source optical emission
behaviour with unprecedented detail.
The analysis of the colour indices reveals that the flux variability can be interpreted
in terms of two components: longer-term variations occurring on a few-day time scale
appear as mildly-chromatic events, while a strong bluer-when-brighter chromatism
characterizes very fast (intraday) flares. By decoupling the two components,
we quantify the degree of chromatism inferring that longer-term flux
changes imply moving along a $\sim 0.1$ bluer-when-brighter slope in the $B-R$ versus 
$R$ plane; a steeper slope of $\sim 0.4$ would distinguish the shorter-term variations.
This means that, when considering the long-term trend, the $B$-band flux level is related 
to the $R$-band one according to a power law of index $\sim 1.1$.
Doppler factor variations on a ``convex" spectrum could be the mechanism accounting for both
the long-term variations and their slight chromatism.

\bigskip]

\section{Introduction}

Blazars are a class of radio-loud active galactic nuclei (AGNs)
well known for their peculiar properties,
such as their intense and extremely variable non-thermal emission across all the 
electromagnetic spectrum from the radio band to $\gamma$-rays (sometimes up to
TeV energies).
Blazars are divided into two subclasses: flat-spectrum radio quasars and BL Lac
objects,
whose main difference is the lack or weakness of emission lines in BL Lac
objects.

The commonly accepted paradigm capable of explaining most of the observed 
radio-loud AGN properties involves a supermassive
black hole surrounded by an accretion disc, feeding a relativistic plasma jet
emitting synchrotron and inverse-Compton radiation.
In this scenario blazars would be the fraction of sources with the jet
oriented at a small angle with respect to the line of sight.

To investigate the details of the physical processes and geometric 
conditions
at the base of the extreme emission variability, a great observing effort is 
needed, possibly organizing multiwavelength campaigns to
follow the source emission behaviour simultaneously at different frequencies.
For example, the Whole Earth Blazar Telescope 
(WEBT; {\tt http://www.to.astro.it/blazars/webt/}; 
e.g.\ Villata et al.\ \cite{vil00}, \cite{vil02})
is an international collaboration of optical and radio observers whose aim is 
the intensive 
and accurate monitoring of selected blazars during time-limited campaigns.
The distribution in longitude of the WEBT members makes continuous (24 hours per day)
optical monitoring possible, at least in principle.
Since its inception in 1997, various BL Lac objects have been the target of the 
WEBT\footnote{Targets of WEBT campaigns have been, besides BL Lacertae, 
\object{AO 0235+16} (Raiteri et al.\ \cite{rai01}),
\object{Mkn 421}, \object{S5 0716+71} (Villata et al.\ \cite{vil00}), 
\object{Mkn 501}, \object{3C 66A}.};
in particular, four campaigns have been devoted to \object{BL Lacertae}, 
the prototype of the above mentioned subclass.

The first two BL Lac campaigns were organized in 1999 in conjunction with pointings
of the X-ray satellites BeppoSAX (June 5--7) and ASCA (June 28--30).
They lasted a few days only and involved a restricted participation of the WEBT 
members.
The results of the BeppoSAX-WEBT campaign have been published in Ravasio et al.\
(\cite{rav02}): the shape of the X-ray spectrum was concave, with a very hard 
component 
above 5--6 keV; moreover, a very fast variability event was detected at the 
lowest X-ray energies, with no simultaneous optical counterpart.

The subsequent WEBT BL Lac 2000 campaign was an impressive observing effort with
a long extension in time: more than 15000 optical observations were performed by 24 
telescopes in 11 countries from May 2000 to January 2001. 
The  results of these observations are reported in Villata et al.\ 
(\cite{vil02}). The original motivation for the campaign
was to provide the low-energy observing counterpart for a high-energy campaign 
organized in July--August 2000, involving X-ray satellites as well as TeV 
detectors (B\"ottcher et al.\ \cite{boe03}).
In this period (the ``core campaign") the densest sampling was achieved,
with observing gaps limited to a few hours, mainly due to the lack of observers 
in the Pacific area.

After the BL Lac Campaign 2000, the WEBT collaboration launched a new
BL Lac campaign in June 2001 (due to the detection of a
fast brightening of the source), which lasted until the end of February
2002. In this paper we present the optical light curves of this BL Lac 2001
campaign and of its so-called ``extension", which includes the two WEBT campaigns
of 1999 and a collection of unpublished data back to 1994.
We also include in our analysis previously published
optical data from the BL Lac Campaign 2000 (Villata et al.\ \cite{vil02}) and 
literature data from 1994 to 2002. As a result, the optical
light curves that we present in this work, based on new and previously
published WEBT and literature data, cover a period of more than 8 years
(9 observing seasons).
Radio and optical light curves covering the period 1968--2003 will be presented 
in a forthcoming paper (Villata et al.\ \cite{vil04}),
where timing and cross-correlation analyses will be performed on both
optical and radio light curves.

In Sect.\ 2 we report the optical observations of BL Lacertae carried out
by the WEBT collaboration; the $UBVRI$ light curves from 1994 to 2002 are
presented in Sect.\ 3, and a detailed colour analysis is performed in Sect.\ 4.
Conclusions are drawn in Sect.\ 5.

\begin{table*}
\caption{List of participating observatories by longitude: $d$ is the telescope diameter;
$N_{\rm obs}$ is the total number of observations done, i.e.\ the number of
unbinned data; $N_U$, $N_B$, $N_V$, $N_R$, and $N_I$ are the
numbers of data points in $UBVRI$ remained after discarding and binning some of
the original data; the numbers in brackets refer to the WEBT campaign 2001 alone
(May 2001 -- February 2002).
}
\begin{tabular}{lccccccc}
\hline
Observatory & $d$ (cm)& $N_{\rm obs}$ & $N_U$ & $N_B$
& $N_V$ & $N_R$ & $N_I$ \\
\hline
Kyoto, Japan & 25 & 190 (0) & 0 & 0 & 0 & 19 (0) & 0 \\
Yunnan, China &100 & 25 (25) & 0 &2 (2)&2 (2) &7 (7) & 0 \\
Mt. Maidanak (T-60), Uzbekistan &60&50 (50) &15 (15) &16 (16) &16 (16) &0 &0 \\
Mt. Maidanak (AZT-14), Uzbekistan &50 &150 (150) &49 (49) &47 (47) &49 (49) &3
(3) &0\\
Mt. Maidanak (AZT-22), Uzbekistan &150 &418 (418) &50 (50) &87 (87) &65 (65)
&108 (108) &61 (61)\\
Abastumani, Georgia (FSU) &70 &1823 (316) &0 &276 (36) &232 (0) &903 (213) &150
(0)\\
Crimean, Ukraine &70 &267 (267) &0 &67 (67) &67 (67) &67 (67) &66 (66) \\
Skinakas, Crete &130 &510 (239) &0 &186 (119) &69 (0) &184 (118) &67 (0)\\
Tuorla, Finland &103 &80 (0) &0 &0 &79 (0) &0 &0\\
MonteBoo, Czech Republic &62 &754 (754) &0 &0 &0 &282 (282) &0\\
Vallinfreda, Italy &50 &98 (0) &0 &0 &49 (0) &0 &0\\
Perugia, Italy &40 &524 (419) &0 &0 &112 (112) &208 (180) &166 (125)\\
Greve, Italy &32 &55 (0) &0 &0 &0 &0 &55 (0)\\
Torino, Italy &105 &487 (54) &0 &104 (1) &70 (2) &271 (24) &12 (2)\\
Sabadell, Spain &50 &32 (32) &0 &0 &3 (3) &12 (12) &0\\
Guadarrama, Spain &20 &3 (3) &0 &0 &1 (1) &1 (1) &0\\
Roque (KVA), La Palma &60 &322 (0) &0 &0 &0 &152 (0) &0\\
Roque (KVA), La Palma &35 &42 (42) &0 &0 &10 (10) &10 (10) &0\\
Roque (JKT), La Palma &100 &64 (64) &0 &9 (9) &10 (10) &32 (32) &6 (6)\\
Roque (NOT), La Palma &256 &44 (44) &4 (4) &7 (7) &7 (7) &7 (7) &3 (3)\\
Roque (TNG), La Palma &358 &138 (0) &3 (0) &62 (0) &5 (0) &62 (0) &2 (0)\\
Olin, Connecticut &51 &400 (361) &0 &0 &8 (0) &78 (64) &85 (71)\\
Hopkins, Massachusetts &60 &12 (0) &0 &0 &0 &0 &11 (0)\\
Foggy Bottom, New York &40 &248 (83) &0 &0 &0 &248 (83) &0\\
Boltwood, Canada &40 &96 (96) &0 &6 (6) &6 (6) &6 (6) &5 (5)\\
Bell, Kentucky &60 &23 (23) &0 &0 &0 &22 (22) &0\\
St. Louis, Missouri &36 &423 (340) &0 &10 (4) &40 (16) &161 (137) &39 (15)\\
Sommers-Bausch, Colorado &60 &9 (9) &0 &0 &4 (4) &4 (4) &0\\
Lowell, Arizona &180 &36 (0) &0 &0 &18 (0) &0 &18 (0)\\
San Pedro Martir, Mexico &210 &9 (9) &0 &3 (3) &3 (3) &3 (3) &0\\
San Pedro Martir, Mexico &150 &123 (123) &4 (4) &12 (12) &11 (11) &11 (11) &4
(4)\\
\hline
Total & &   7455 (3921) &125 (122) &894 (416) &936 (384) &2861 (1394) &750 
(358)\\
\hline
\end{tabular}
\end{table*}

   \begin{figure*}
   \centering
   \vspace{2cm}
   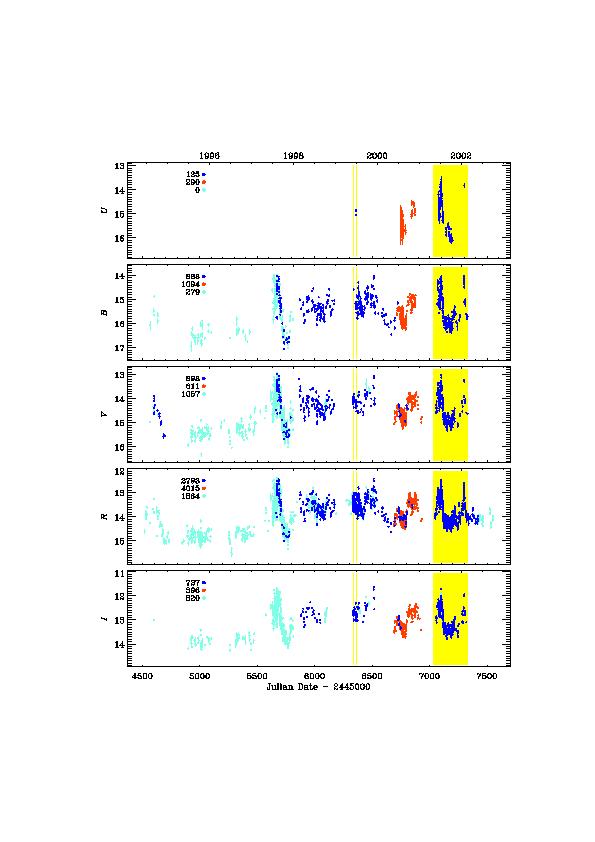
   \vspace{2cm}
   \caption{$UBVRI$ light curves of BL Lacertae from 1994 to 2002;
   the numbers on the left indicate the number of data points in the extended BL Lac
   Campaign 2001
   (blue/dark), in the BL Lac Campaign 2000 (red/grey), and from the literature
   (light cyan/light grey);
   yellow (shaded) strips correspond to the periods of the 1999 and 2001 WEBT campaigns.}
   \label{ubvri}
   \end{figure*}

   \begin{figure*}
   \centering
   \vspace{2cm}
   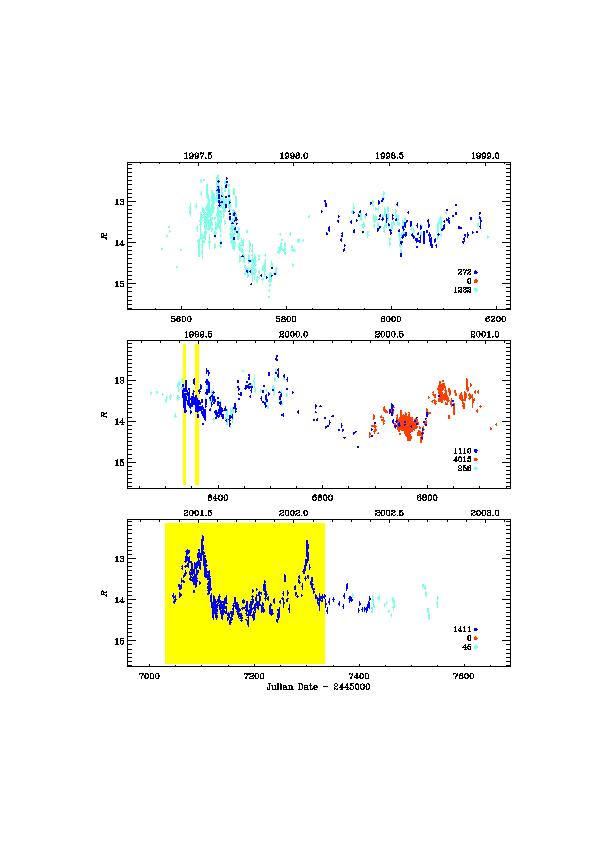
   \vspace{2cm}
   \caption{Folded $R$-band light curve of BL Lacertae from 1997 to 2002;
    symbols as in Fig.\ \ref{ubvri}.}
   \label{rtot_fold}
   \end{figure*}

   \begin{figure*}
   \centering
   \includegraphics{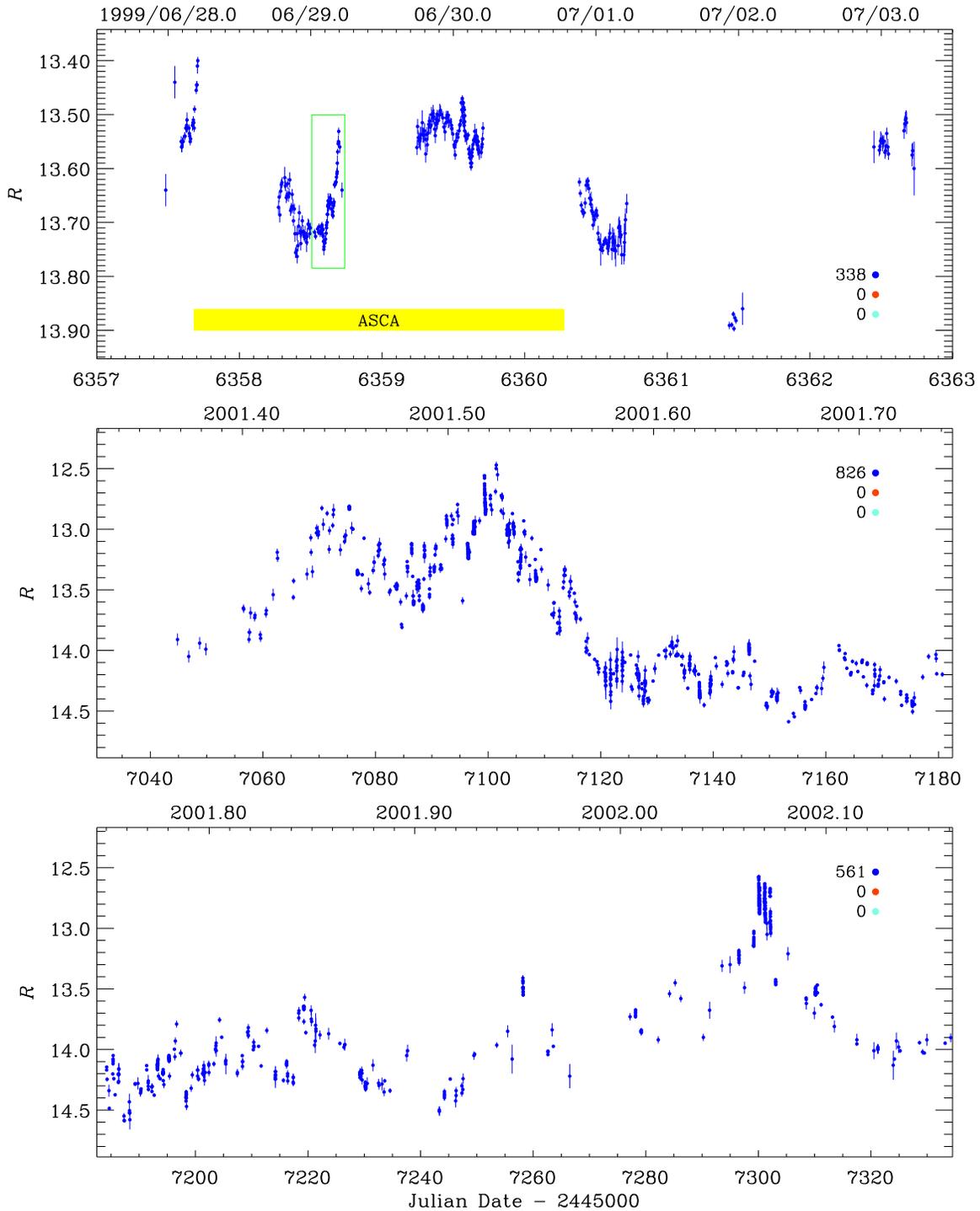}
   \caption{Enlargements of the BL Lacertae $R$-band light curve corresponding
    to the
    WEBT campaigns of late June -- early July 1999 (top panel) and May 2001 --
    February 2002
    (middle and bottom panels); symbols as in Fig.\ \ref{ubvri}; in the top
    panel the yellow (grey) strip indicates the period of the ASCA pointing.}
   \label{r_camp}
   \end{figure*}

   \begin{figure*}
   \centering
   \includegraphics[width=13cm]{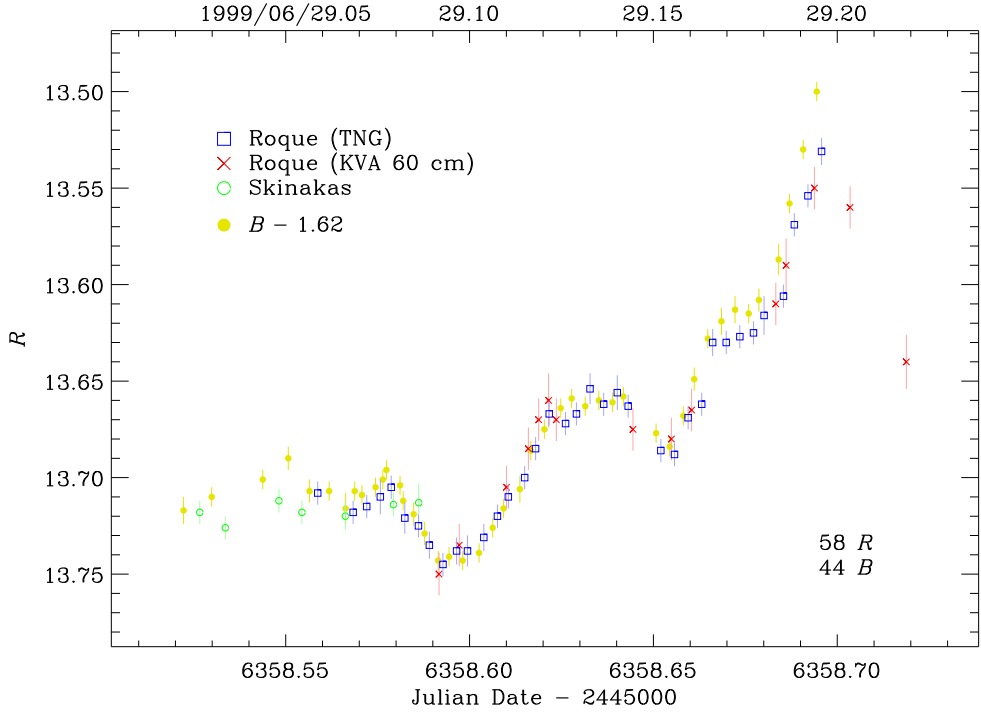}
   \caption{\ An enlargement of the $R$-band light curve of BL Lacertae on June 29, 1999,
    corresponding to the green (grey) box in Fig.\ \ref{r_camp}; the
    yellow (grey) filled circles represent the simultaneous $B$-band light curve 
    shifted by $-1.62$ mag.}
   \label{rtot_pap2}
   \end{figure*}

   \begin{figure*}
   \centering
   \vspace{2cm}
   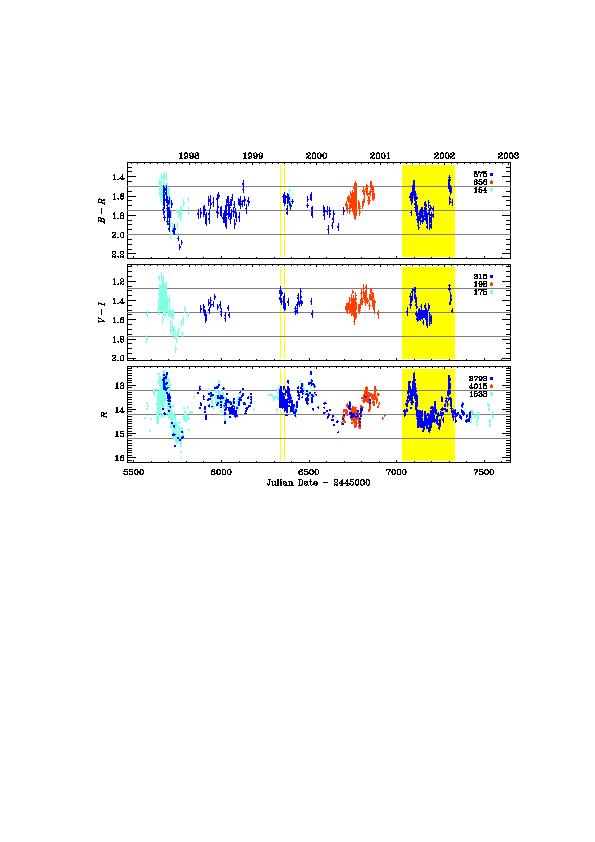
   \vspace{2cm}
   \caption{$B-R$ (top panel) and $V-I$ (middle panel) colour indices
    versus time from 1997 to 2002 compared to the
    $R$-band light curve; the host-galaxy contribution has been removed from the data;
    symbols as in Fig.\ \ref{ubvri}; the horizontal lines divide each panel into four equal 
    ranges (1 mag for the light curve and 0.25 mag for the colour indices) for clarity.}
   \label{colori_jd}
   \end{figure*}

   \begin{figure*}
   \centering
   \includegraphics{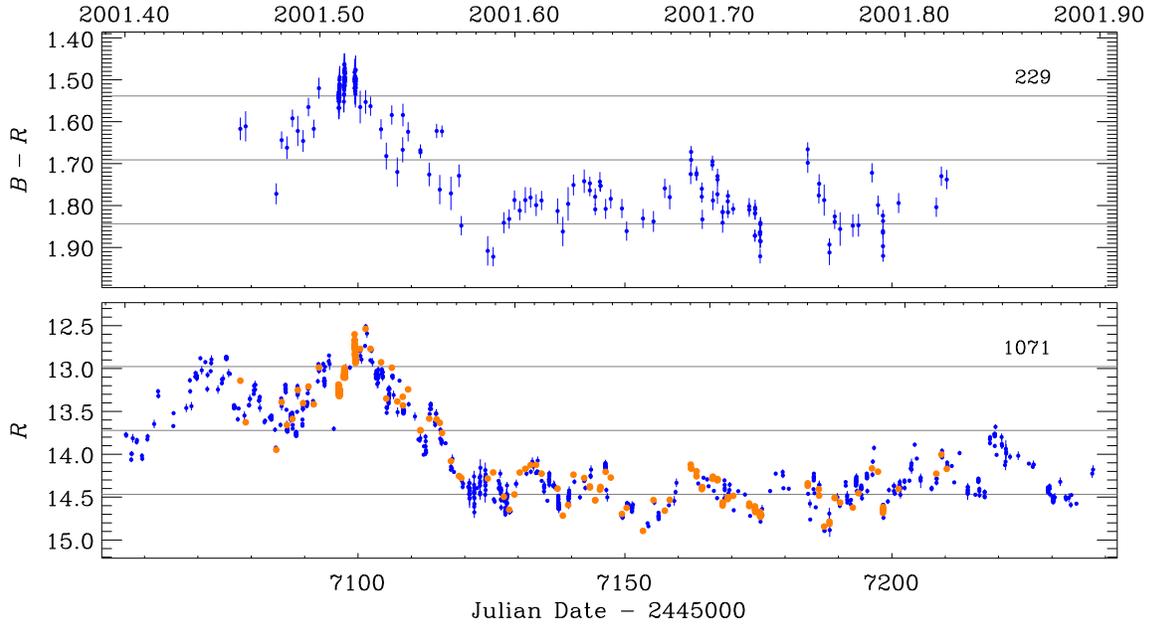}
   \caption{An enlargement of the $B-R$ versus time curve (top panel) compared
    with the $R$-band light curve (bottom panel) during the best-sampled period of the 
    BL Lac Campaign
    2001; orange (grey) dots represent the data used for calculating the colour indices.}
   \label{colori_jd2}
   \end{figure*}

   \begin{figure*}
   \centering
   \includegraphics[width=13cm]{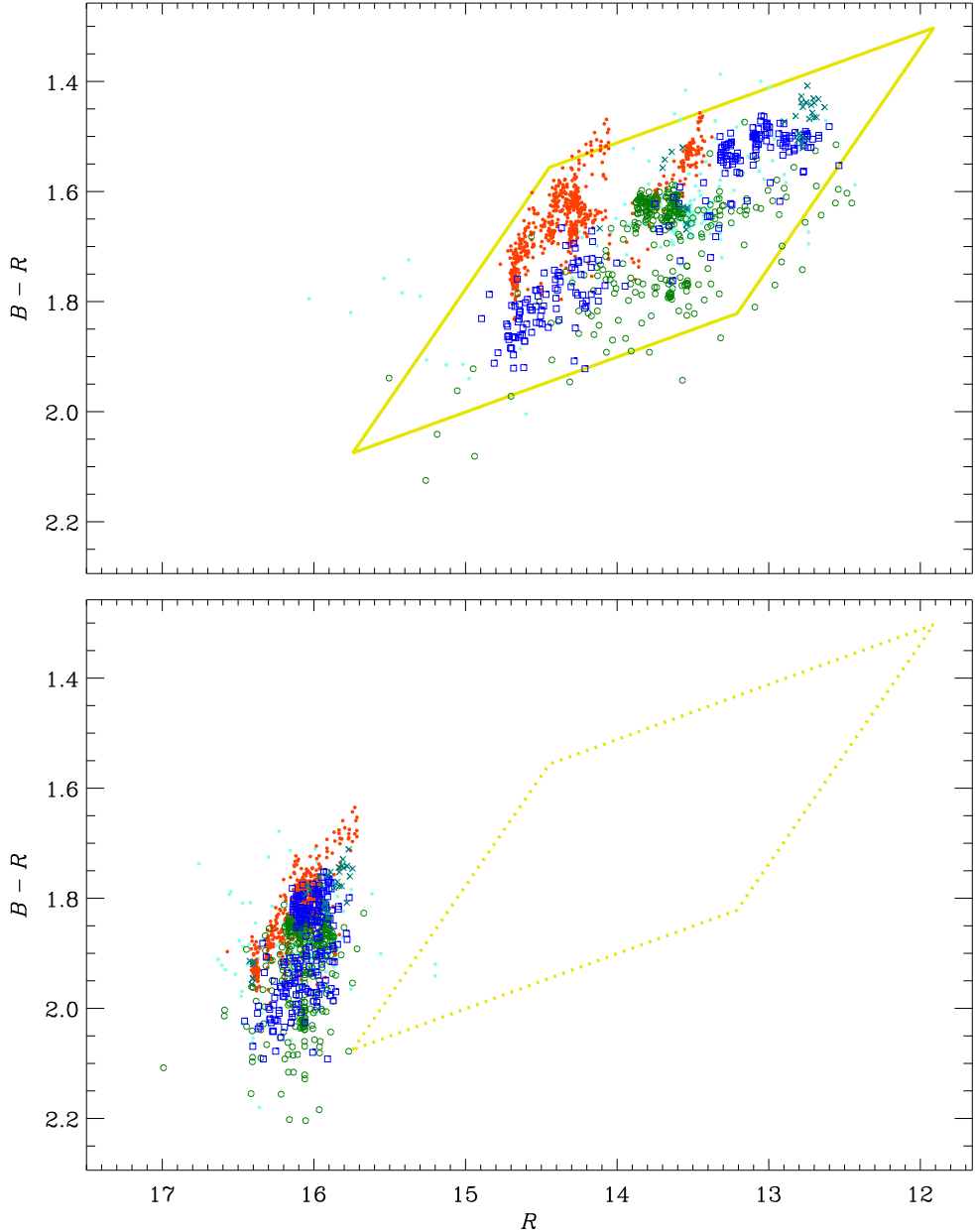}
   \caption{\ $B-R$ colour indices versus $R$ magnitudes after the host-galaxy
    removal (upper panel):
    BL Lac 2000 and literature data are represented by red (grey) and light
    cyan (light grey) dots,
    respectively, while BL Lac 2001 extended data are plotted with green (grey)
circles (BL Lac 2001 extension), blue (dark) squares (best-sampled period of BL Lac
2001, the same of Fig.\ \ref{colori_jd2}), and cyan (grey) crosses (remaining part of BL
Lac 2001); most of the data lie within a parallelogram whose sides have slopes of 0.1
and 0.4. In the bottom panel the same data have been corrected for the long-term
trend as explained in the text.}
   \label{b-r_vs_r}
   \end{figure*}

   \begin{figure*}
   \centering
   \vspace{2cm}
   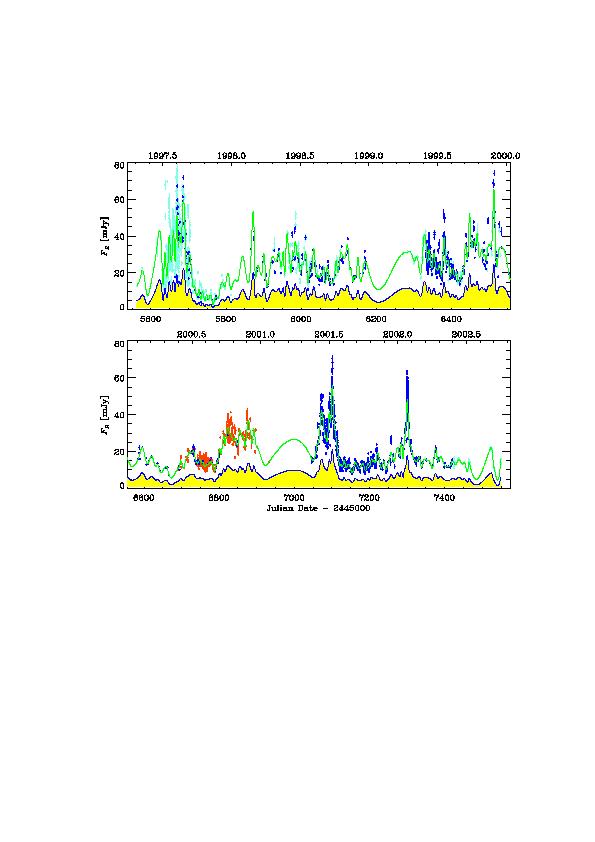
   \vspace{2cm}
   \caption{Galaxy-subtracted $R$-band flux light curve from 1997 to 2002
    (folded, symbols as in Fig.\ \ref{ubvri});
    the green (grey) line represents a cubic spline interpolation through the
    4-day binned
    light curve; the blue (dark) line, obtained by normalizing the spline to its
    minimum value,
    represents the actual correction factor for removing the long-term trend.}
   \label{fitta_fold}
   \end{figure*}

\section{Observations of BL Lacertae by the WEBT collaboration}

The BL Lac Campaign 2001 was started in June 2001 after the detection
of a fast brightening of the source.
Optical and radio monitoring was continued until the end of February 2002.
However, at the time of data analysis, it was decided to collect all optical
and radio data taken by the members of the WEBT collaboration dating back to 1968.
The observations performed out of the proper campaign period (May 2001 -- February 2002)
constitute the so-called ``campaign extension"; they also include the still 
unpublished data of the WEBT campaign of late June 1999 and the partially
published (Ravasio et al.\ \cite{rav02}) data of the early-June 1999 campaign.
In this first paper, we present the optical data only, starting from 1994.

Table 1 shows the list of optical observatories that participated 
in the data collection
with unpublished data or data from the early-June 1999 campaign.
The name and location of the observatory (Col.\ 1) is followed
by the telescope diameter (Col.\ 2), by the total number of observations made
(Col.\ 3) and the number of data points in $UBVRI$ derived,
sometimes after binning (Cols.\ 4--8).
The numbers in brackets refer to the WEBT campaign 2001 alone.

The observing strategy during the BL Lac Campaign 2001 was nearly the same
as in the BL Lac Campaign 2000 (Villata et al.\ \cite{vil02}): in general,
observers with telescopes larger than 60 cm were invited to perform
$BR$ sequences of frames during each observing night, along with a complete
($U$)$BVRI$ sequence at the beginning and at the end.
Exposure times were chosen to obtain a good compromise between high
precision and high temporal density. Participants with smaller-size telescopes
were suggested to carry out observations in the $R$ band only.

In general, CCD data were collected as instrumental magnitudes of
the source and reference stars to apply the same analysis and
calibration procedures to all datasets. Instrumental magnitudes were obtained
by either standard aperture photometry
procedures or Gaussian fitting, in most cases by the observers themselves.
A minority of data came from photometer observations: these data were provided
directly as standard magnitudes.

A careful data assembly was required because of the huge amount of data coming
from many different instruments: details can be found in Villata et al.\
(\cite{vil02}).

\section{Optical light curves}

$UBVRI$ light curves from 1994 to 2002 are shown in Fig.\ \ref{ubvri}.
Blue (dark) dots refer to data belonging to the BL Lac Campaign 2001 plus
extension (see Sect.\
2); red (grey) points represent the BL Lac Campaign 2000 data
(Villata et al.\ \cite{vil02}); literature data\footnote{Literature data
have been taken from Maesano et al.\ (\cite{mae97}),
Fan et al.\ (\cite{fan98}; with dates taken from Fan \& Lin \cite{fan00}), Webb
et al.\ (\cite{web98}),
Bai et al.\ (\cite{bai99}), Tosti et al.\ (\cite{tos99}), Ghosh et al.\
(\cite{gho00}), Katajainen et al.\ (\cite{kat00}),
Clements \& Carini (\cite{cle01}), Fan et al.\ (\cite{fan01}), Sobrito et al.\
(\cite{sob01}),
Papadakis et al.\ (\cite{pap03}), Tosti et al.\ (\cite{tos04}). Sometimes the data
used in this paper
do not  exactly correspond to the original ones since some correction/removal
(often done in agreement with the data owners) was needed. Moreover, some datasets
were completely re-analysed.} are shown as light cyan (light grey) dots.
The periods of the 1999 and 2001 WEBT campaigns on BL Lacertae are highlighted
by yellow (shaded) strips.

Each panel in the figure covers a range of 4 magnitudes. One can easily see that the
variation amplitude is greater at higher frequencies, which is a common 
behaviour in blazars. For instance, during the well-sampled 1997 outburst, 
the brightness excursion was 3.09 mag in the $B$ band, 3.06 mag in $V$, 2.94 mag in $R$, 
and 2.47 mag in $I$. Hence, the variation in the $B$ band was 5\% wider than 
in the $R$ band and 25\% wider than
in the $I$ band. Actually, these values of the $\Delta B$ excess have to be intended
as underestimates, because the $B$-band light curve is undersampled,
in particular around maxima and minima. In general, in this period the
number of $B$ data is only 18\% of the $R$ one and 44\% with respect to the $I$ data.

Another example can be derived by considering the period of the WEBT campaign 2001.
Total magnitude variations were $\Delta U = 2.63$, $\Delta B = 2.34$,
$\Delta V = 2.32$, $\Delta R = 2.12$, and $\Delta I = 2.07$.
Thus, again in spite of their poorer sampling (see Table 1),
the $UBV$ data show a larger total variation with respect to  the $R$ (and $I$) band.
   
From Fig.\ \ref{ubvri} one can also notice that, while the minimum brightness
states are very different, there seems to be a well-defined brightness upper limit.
Indeed, the brightest levels detected in summer 1997, late 1999, mid 2001
and early 2002 reached about the same value.

The best-sampled light curve is the $R$-band one (8672 data points); in Fig.\
\ref{rtot_fold} we show the most interesting period, 1997--2002, folded in
three panels.
Each panel covers two observing seasons, starting and ending on February 17 
(solar conjunction).
The most noticeable features are the 1997 outburst, well covered by literature 
data, and the other
two strong outbursts occurring in mid 2001 and early 2002, during the BL Lac 2001 
WEBT campaign.
Other well-sampled periods are mid 1999 (especially the WEBT campaigns) and of 
course the BL Lac
Campaign 2000. The 1997 outburst still remains the most spectacular event for 
both its magnitude 
excursion during the dimming phase and the large amplitude of the
short-term variations (see e.g.\ Villata et al.\ \cite{vil02} for references).

The three panels in Fig.\ \ref{r_camp} display the details of the WEBT campaigns
   carried out in mid 1999 contemporaneously with the ASCA pointing (top)
   and in 2001--2002 (middle and bottom); they
   correspond to the second and third yellow (shaded) areas of Fig.\
   \ref{rtot_fold}.
   Details of the 1999 WEBT-BeppoSAX campaign (first yellow/shaded strip in Fig.\
   \ref{rtot_fold})
   have already been presented in Ravasio et al.\ (\cite{rav02}).
   In the upper panel of Fig.\ \ref{r_camp}, spanning the period June 27.5 -- July 3.5, 1999,
   the time coverage by the ASCA satellite is indicated by a yellow (grey) strip;
   a comparison
   between the optical and X-ray data will be performed elsewhere.
   A very well-sampled intranight variation was observed
   at the beginning of June 29 (green/grey box), which is better displayed in Fig.\
   \ref{rtot_pap2}, where
   both the $R$-band and the $B$-band data are shown to
reveal the expected larger variation in the $B$ band:
$\Delta B=0.243$ versus $\Delta R=0.219$ in 2.5 h.
Notice how data from different telescopes are in fair agreement and
how the well-defined trend confirmed by high-precision data in two bands 
provides the finest details of the variation.
   
   In the 2001--2002 WEBT campaign, one can notice the double-peaked,
   nearly symmetric outburst of mid 2001, with fast oscillations superimposed
   (Fig.\ \ref{r_camp}, middle panel). On the contrary, the outburst occurring at
   the beginning of 2002 appears as a peculiar event with respect to the previous ones
   because of its single, short-duration peak.
   However, one can recognize fast (intranight) flares 
   perturbing the main trend (Fig.\ \ref{r_camp}, bottom panel).

\section{Colour analysis}

As done for the BL Lac Campaign 2000 (Villata et al.\ \cite{vil02}), the 
first step of the colour analysis
is the subtraction of the host-galaxy contribution from the fluxes to avoid
its colour contamination.
We have estimated the host-galaxy magnitudes in the various bands using data 
published by
Scarpa et al.\ (\cite{sca00}) and Mannucci et al.\ (\cite{man01}), obtaining
$U_{\rm host}=17.65$, $B_{\rm host}=17.15$, $V_{\rm host}=16.16$, $R_{\rm 
host}=15.55$, and $I_{\rm host}=14.92$.
Conversion of magnitudes into fluxes has been done as in Villata et al.\ 
(\cite{vil02}), where it was also estimated that
only about 60\% of the host-galaxy flux must be subtracted from the original
data.

Once this procedure was performed, in BL Lac 2000 the resulting colour indices 
showed a fairly clear behaviour:
long-term variations appeared essentially achromatic, while fast flares 
exhibited a strong bluer-when-brighter trend.

In Fig.\ \ref{colori_jd} we show the time evolution of the $B-R$ and $V-I$ BL 
Lac colour indices from 1997 to 2002, 
compared with the galaxy-subtracted $R$-band light curve. We have now a dataset 
which is much more extended 
both in time and in brightness. Indeed, the outburst of BL Lac 2000 appears as a 
modest event when compared to
the 1997, 2001, and 2002 outbursts. It is evident that these main outbursts show
a chromatic behaviour, even if the colour changes observed during the short-term
flares are relatively stronger.
By looking at the 2001 outburst one can see that the total range
covered by the $B-R$ index is about
twice the range covered during fast flares, whereas the ratio between the
corresponding magnitude variations is about three.
This situation is better displayed in Fig. \ref{colori_jd2}.

   The long-term achromatism found by Villata et al.\ (\cite{vil02}) in BL Lac 2000
   appears to be due to a bluer-than-usual pre-outburst state,
   which we can now recognize as a peculiar event.
   Hence, from now on we will speak of ``strongly chromatic" fast flares and
   ``mildly chromatic" long-term variations.
   
We now investigate whether the global behaviour of the source 
can be uniformly described
in terms of these two components. We try to decouple the two components in
a way similar to that followed by Villata et al.\ (\cite{vil02}).
   
In the upper panel of Fig.\ \ref{b-r_vs_r} we have plotted the $B-R$ colour 
indices versus
the galaxy-removed $R$ magnitudes. Red (grey) dots represent the data of BL Lac 2000,
literature data are displayed as light cyan (light grey) dots, and data from
the BL Lac 2001
campaign plus extension are plotted with different symbols according to
the observing period (see caption for details). One can see the different 
behaviour during the two
campaigns: while $B-R$ data of BL Lac 2000 are located along two separated 
quasi-linear trends with
similar slopes and colour ranges (thus reflecting the quasi-achromatic long-term behaviour), 
the other ones are more 
uniformly distributed. As a whole, the index dataset is mostly comprised
within a parallelogram with slopes 0.1 and 0.4. 

   \begin{figure*}
   \centering
   \vspace{2cm}
   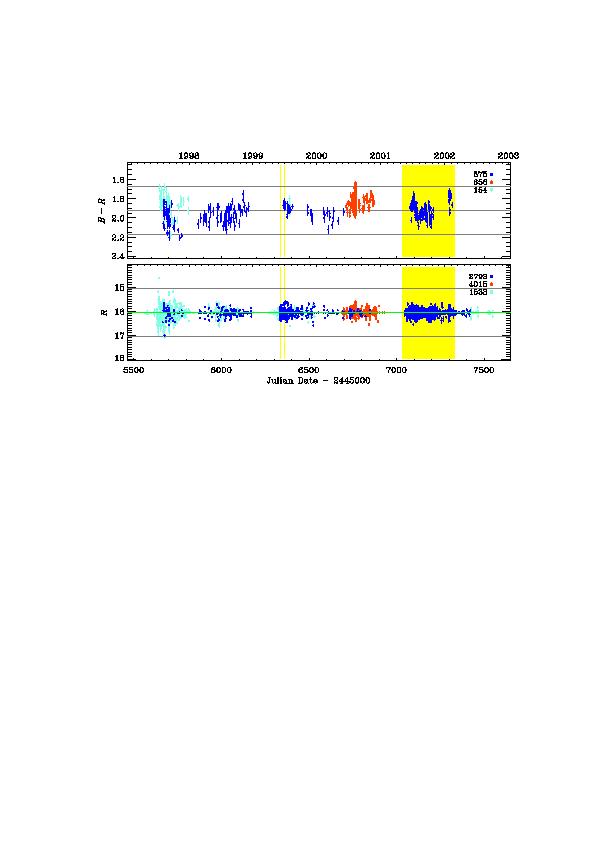
   \vspace{2cm}
   \caption{$B-R$ and $R$-mag curves as obtained after correcting for
    the long-term trend (symbols as in Fig.\ \ref{ubvri}, horizontal lines as 
    in Fig.\ \ref{colori_jd}); the green (grey) line at $R = 16.061$ in the bottom panel
    represents the minimum level of the spline (see Fig.\ \ref{fitta_fold})
    to which the light curve has been normalized.}
   \label{colori_jd_cor}
   \end{figure*}

A 0.1 slope trend means a mildly chromatic behaviour that we have recognized
to be typical of the long-term variations. On the contrary, a 0.4
slope trend implies strongly chromatic variations, as found for the fast 
flares. The BL Lac 2000 points near the upper-left corner of 
the parallelogram,
which represent data from the pre-outburst, are essentially aligned with the 
steeper side, but
their colour appears to be bluer than average. Thus, apart from this
discrepancy, we are presenting a scenario where
fast flares imply moving along the steeper direction in the $B-R$ versus $R$ plot,
the position inside the parallelogram depending on the source brightness state.
In other words: 
$$B-R={\rm constant} + 0.4 (R-R_{\rm spl}) + 0.1 R_{\rm spl} \, ,$$
where $R_{\rm spl}$ is the brightness level of the long-term component.

The steeper 0.4 slope has been inferred from
the very well-sampled data of the last week of the core campaign 2000, where
we can assume that the long-term component does not change significantly.
Indeed, by looking at Figs.\ 7 and 8 of Villata et al.\ (\cite{vil02}),
one can estimate
${\rm d}(B-R)/{\rm d} R \approx 0.4$, and a linear correlation analysis
gives a slope of $0.383 \pm 0.004$.

The value of the 0.1 slope has been determined as illustrated in the next subsection.

\subsection{Removal of the long-term trend}

As done in BL Lac 2000, we performed a cubic spline
interpolation across the binned $R$-band flux light curve. To have a good
interpolation through very fast and large variations, like those occurring during
the 2001 and 2002 outbursts, we adopted a 4-day binning.
The resulting spline is plotted as a green (grey) line in Fig.\ \ref{fitta_fold},
showing the folded, galaxy-subtracted flux light curve.

To remove the long-term trend from the light curve, we have rescaled
each original flux by dividing it by the ratio between the value of the spline
at the considered time and its minimum value, $C_R(t)=[F_{\rm spl}(t)/F_{\rm min}]_R$
(where $R$ stands for $R$ band).
This correcting factor is shown in Fig.\ \ref{fitta_fold} as a blue (dark) line 
highlighted by the yellow (shaded) area.

Because of the assumption of an achromatic long-term trend, in BL Lac 2000 the
$B$-band flux light curve was rescaled by the same time-dependent factor. Now, 
the correcting factor for the $B$ fluxes must be found.
Instead of tracing a cubic spline interpolation
across the $B$ fluxes, which would be affected by a different and worse sampling, 
we search for the best-fit relationship between the long-term trends
in the two bands. Then we suppose that $C_B(t)$ is some power of $C_R(t)$.
Various best-fit procedures on different data sub-samples have been applied to find
the value of the power index; this value has been found to range between 1.08 and 1.12.
We adopted 1.1 which, passing from fluxes to magnitudes, explains the 0.1 slope
of the $B-R$ versus $R$ plot.
   
Once both the $R$ and $B$ fluxes have been corrected for their long-term trends,
the resulting colour indices are shown in the bottom panel of Fig.\ \ref{b-r_vs_r}.
What we expect to see in this plot is the signature of the fast flares alone,
shifted to the minimum brightness level of the long-term trend ($R=16.061$).
The final result is fairly satisfactory, but one can still see some inhomogeneous 
behaviour, like the expected bluer level of the BL Lac 2000 pre-outburst data.
A further insight into this result is given by Fig.\ \ref{colori_jd_cor}, where
the residual $R$ magnitudes (bottom panel) and corresponding $B-R$ colour 
indices (upper panel) have been plotted versus time. 
By comparing this figure with Fig.\ \ref{colori_jd}, one can see that
the outburst signatures have been completely smoothed away in the $R$-band
light curve, as expected, and almost completely smoothed away in the $B-R$
plot, with the main exception of the bluer-than-average BL Lac 2000 period.
Moreover, one can also notice that the variability range of the fast flares 
in the bottom panel is fairly constant around 1 mag, but it seems to exceed 1.5 mag 
or more during the 1997 outburst. This represents another peculiarity of an outburst
with respect to the others. Part of the amplitude excess may be explainable by 
inhomogeneity of the literature datasets, but not all of it.

\section{Conclusions}

We have presented $UBVRI$ light curves of BL Lacertae from 1994 to 2002 composed
from data taken by the members of the WEBT collaboration as well as literature data.

The colour analysis we performed on the best-sampled period of these light curves
(1997--2002) leads to the following results.

The variability observed in the optical light curves can be interpreted in terms
of two components: a ``mildly-chromatic" longer-term
component and a ``strongly-chromatic" shorter-term one.
The longer-term component has a typical time scale of a few days,
while the shorter-term one is responsible for the very fast, intraday flares, and
determines a steep bluer-when-brighter slope of about 0.4 in the $B-R$ versus $R$ plot.
Moreover, our analysis suggests that, in the long-term, the $B$-band flux level is 
related to the $R$-band one according to a power law with index 1.1. 
This means that long-term variations trace a 0.1 slope in the $B-R$ versus $R$ plot.
In other words, all possible flux variations fill
a parallelogram with 0.1 and 0.4 slopes in the $B-R$ versus $R$ plane.

The analysis presented in this paper shows that, in general,
the BL Lac optical behaviour can be uniformly described in terms of these two components
fairly well, with the caution that the presented decoupling method can work 
well only during well-sampled periods, 
where the light-curve binning can actually represent the long-term trend.
However, peculiar events also occurred, as
in the BL Lac 2000 period, and our scenario seems to be more a good compromise between
slightly different behaviours rather than a definitive solution. 
Other factors appear to change in the source history, in one or both of the components.

Changes in the flaring component could explain the 
larger amplitudes of the 1997 outburst and the bluer BL Lac 2000 period, 
but the remaining blue residues of the 1997, 2001, and 2002 
outbursts suggest other possibilities.

In Villata et al.\ (\cite{vil02}) the quasi-achromatism of the long-term variations 
was interpreted in terms of Doppler factor variations. 
If the intrinsic source spectrum around the infrared--optical bands is well 
described by
a power law ($F_\nu \propto \nu^{-\alpha}$), a Doppler factor variation does not imply a
colour change. What about a mildly-chromatic behaviour? It could be due to a Doppler
factor variation on a spectrum slightly deviating from a power law (convex for 
a bluer-when-brighter trend). If the spectrum 
shape changes in time, the chromatism of the outbursts also varies, and the 0.1 $B-R$
slope adopted in this paper might represent only a mean value among different
slopes. In other words, any period should be treated with a proper slope. 
A further investigation of this is beyond the scope of this paper.

The strongly-chromatic fast flares are likely due to intrinsic phenomena, such as
particle acceleration from shock-in-jet events, widely described in the literature 
(see e.g.\ Mastichiadis \& Kirk \cite{mas02} for a review). Other models are also
discussed in the recent paper by B\"ottcher \& Reimer (\cite{boe04}).

Thus, the present extension/enrichment of the BL Lac optical dataset 
partially confirms previous results leading to a more refined scenario, 
but also gives rise to new intriguing questions.

\begin{acknowledgements}
This research has made use of
the NASA/IPAC Extragalactic Database (NED), which is operated by the
Jet Propulsion Laboratory, California Institute of Technology, under
contract with the National Aeronautics and Space Administration.
This work was partly supported by the European Community's Human Potential Programme
under contract HPRN-CT-2002-00321, by the Italian Space
Agency (ASI) under contract CNR-ASI 1/R/27/02, and by the Italian Ministry for
University and Research (MURST) under grant Cofin 2001/028773.
It is based partly on observations made with the Italian Telescopio Nazionale 
Galileo (TNG) and the Nordic Optical Telescope, both operated on the island of 
La Palma at the Spanish Observatorio del Roque de los Muchachos of the Instituto 
de Astrof\'{\i}sica de Canarias, the former by the Centro Galileo Galilei of the 
INAF (Istituto Nazionale di Astrofisica) and the latter jointly by Denmark, 
Finland, Iceland, Norway and Sweden. St.-Petersburg group was supported by Federal
Programs ``Astronomy" (grant N 40.022.1.1.1001) and ``Integration" (grant N B0029). 

\end{acknowledgements}

\end{document}